\documentstyle[12pt,aps,epsf]{revtex}

\input{epsf}
\begin{document}

\title{Stored Light and Released Fiction}
\author{E.B.Alexandrov and V.S.Zapasskii}
\address{All-Russia Research Center ``Vavilov State Optical Institute'',
 St. Petersburg, 199034 Russia\\
 e-mail: zap@vz4943.spb.edu
 }

\maketitle
\vskip10mm

\begin{abstract}
It is shown that the interpretation of the experimental results
reported in the publication ``Storage of Light in Atomic Vapor'' by
D.F.Phillips et al. $\cite{phil}$ is incorrect.  The experimental
observation of $\cite{phil}$ can be consistently explained in the
framework of standard concepts of the physics of optical pumping
and have nothing to do with ``storage of light'', or ``dynamic
reduction of the group velocity'', or ``light pulse compression''.
\end{abstract}

\section{Introduction}

Our attention was attracted to paper \cite{phil} by our colleague who
decided to use the results of this work to ``trap, store, and
release'' the object beam in a telescopic system.  Being aware of
our experience in magneto-optics and optical pumping in alkali
vapors, he asked us to comment about experimental details of
this publication.  So, we had to thoroughly examine this paper.
The result of this examination was rather unexpected: it became
evident that the authors' claims on the "light storage" have
very little to do with real physical processes in the system.

\section{Experimental}

Recall that, in the work under consideration, rubidium atoms (in
zero magnetic field) were excited by a laser tuned to the
transition $5^2S_{1/2}$, $F=2 \rightarrow 5^2P_{1/2}$, $F^{\prime} =1$.
The authors considered
the exciting field as consisting of two coherent beams with
orthogonal circular polarizations, namely, a strong $\sigma_+$-polarized
``control'' beam and a weaker $\sigma_-$-polarized ``signal'' beam, which
was turned on for a few tens of microseconds.  After passing
through the cell with Rb vapors, the two circularly polarized
beams were detected by separate photodetectors.  It is important
that these two beams were treated independently, i.e., the
$\sigma_+$-component at the exit of the medium was always attributed to
the control beam ($\sigma_+$-polarized at the entrance), while the
$\sigma_-$-component, to the signal beam ($\sigma_-$-polarized at the entrance).
This approach contains a simple and fatal error that makes all
the authors' interpretation inadequate.

As experimentalists, we
will analyze the observations described in \cite{phil} in general terms
on the basis of simplest physical considerations to make the
experimental situation as transparent as possible.  We will
leave aside the vast theoretical background of \cite{phil} for the
reason that we strongly doubt about correctness of the simplest
model (the three-level $\Lambda$-scheme) used by the authors in the
situation when the studied atomic line comprises 16 levels, with
11 of them being directly involved into the game through
radiative and collisional processes.  In addition, the strong
fields used in the experiment connect 8 of these levels, thus
making approximate methods inapplicable.

\section{The picture of the observations}

In fact, the authors examined a response of the atomic system to
a pulsed modulation of the pumping circularly polarized beam,
namely, during a time interval of 10 - 30 $\mu$s the beam was
polarized elliptically.  In the fundamentally nonlinear
experiment (which we deal with), the medium excited by an
elliptically polarized beam, with a distinguished direction in
the plane of the wavefront, loses its axial symmetry (initially
provided by the absence of external magnetic fields and by the
axial circularly polarized pump), the circularly polarized waves
cease to be eigenwaves of the medium, and the authors' approach,
which implies mutual independence of the circularly polarized
waves in the medium, becomes invalid.  Now, the ``control''
$\sigma_+$-polarized beam will detect the elliptic anisotropy of the
vapor.  It means that the beam will exhibit elliptic
birefringence and dichroism (the latter being most important),
and its polarization state will vary upon its propagation
through the medium.  At the exit of the medium, the ``control''
beam will prove to be elliptically (rather than circularly)
polarized and, being decomposed into the basis circular
polarizations, will contribute to the signal of the second
detector which is supposed to detect the ``signal'' $\sigma_-$-polarized
beam.  The same is valid for the ``signal'' beam, which will be
detected by both detectors also.  This is, briefly, what the
authors observe.

\section{Three ``surprising'' stages of the experiment}

In more detail, the mechanism of the observed polarization
dynamics looks as follows (stages 1, 2, and 3 are indicated in
the figure).

{\bf Stage 1}.  The medium, preliminary oriented by the
strong ``control'' beam, is subjected to a pulse of elliptically
polarized excitation, which renders the medium partially
aligned.  In view of relatively small duration of the ``signal''
pulse (as compared with the spin alignment time constant), the
degree of alignment continuously increases during the pulsewidth
(as the integral of the pulse energy), as well as does the
fraction of the ``control'' beam detected in the ``signal'' channel.
This process evidently has nothing in common with the ``reduction
of the light group velocity'' in the medium.  The growth of the
signal in stage 1 reflects not the shape of the leading edge of
the ``signal'' pulse, but rather the process of accumulation of
the spin alignment.  The transparency of the medium for the
circularly polarized beam starts to slowly move toward its
transparency for the light of elliptic polarization.

{\bf Stage 2}.  The light is turned off, and the alignment, thus accumulated or,
more exactly, all the accumulated polarization moments remain
stored in the system, relaxing with a characteristic time of
about 1 ms.  No energy is stored in the system and there are no
grounds to be surprised that no output signal is observed as
long as the light is off.

{\bf Stage 3}. After the dark pause (stage
2), the ``control'' beam goes on detecting the residual elliptic
anisotropy of the medium (partially relaxed during the pause),
which is revealed, as before, in the nonorthogonality of the two
circular polarizations in the medium and in branching of a
fraction of the ``control'' beam into the ``signal'' channel.  Now,
however, the control beam, bringing strong orientation to the
system, destroys the alignment much faster than it occurred in
the darkness.  The shape of the tail observed after the dark
pause has nothing to do with that of the signal pulse.  This
tail reflects dynamics of the alignment destruction under the
action of the control beam. This is why the duration of this
tail will, in particular, depend on the control beam power, and
the authors will never be able to ``store'' any other part of the
pulse than its exponential-like tail.  It may also seem curious
for the authors' interpretation that the total amount of the
light emerging from the cell when the ``stored'' pulse is being
``released'' is smaller than when the ``release'' is over.  This
fact can be easily verified experimentally.

Thus, when we take
into account the fact that the elliptically polarized light
makes the medium elliptically anisotropic, all the observations
became absolutely natural and trivial.

\section{What is really stored}

In a sense, the control beam, turned on after the pause, indeed
reads out the information about the relative phase and energy of
the signal beam.  This information is stored in the magnitude
and azimuth of the alignment and determines the phase and
amplitude of the wave projected to the signal channel.  But,
first, the authors do not measure the phase of the signal wave
and, second, what has it got to do with the "storage of light"?
The sense of this term the authors unambiguously explain in the
fragment: "A pulse of light which is several kilometers in free
space is compressed to a length of a few centimeters and then
converted into spin excitation of a vapor of Rb atoms. After a
controllable time, the process is reversed and the atomic
coherence is converted back into a light pulse".  In our
opinion, the experiment [1] is so simple that it does leave any
room for this kind of speculations.  Note, in addition to all
the aforesaid, that the atomic coherence carries no energy in
the absence of magnetic field: the acquired "spin excitation"
means just a sort of spatially uniform spin anisotropy of the
medium.  As one can easily see, in the authors' terminology, any
quarter-wave plate "stores" circularly polarized light that can
be "released" by irradiating it with circularly polarized light
of opposite handedness.  We do not find this language sensible.

\section{Conclusion}

The interpretation of the results presented in [1] contains
evident errors, and the reported experimental data (which, in
themselves, seem quite correct) provide no evidence for the
light ``trapping, storage, and release''.  In our opinion, this
paper not only misleads the reader, but also arouses suspicion
about other publications pertaining to the "storage of light"
stream.  This why we considered adequate evaluation of this
paper important.

\begin{figure}
\epsfxsize=400pt
\epsffile{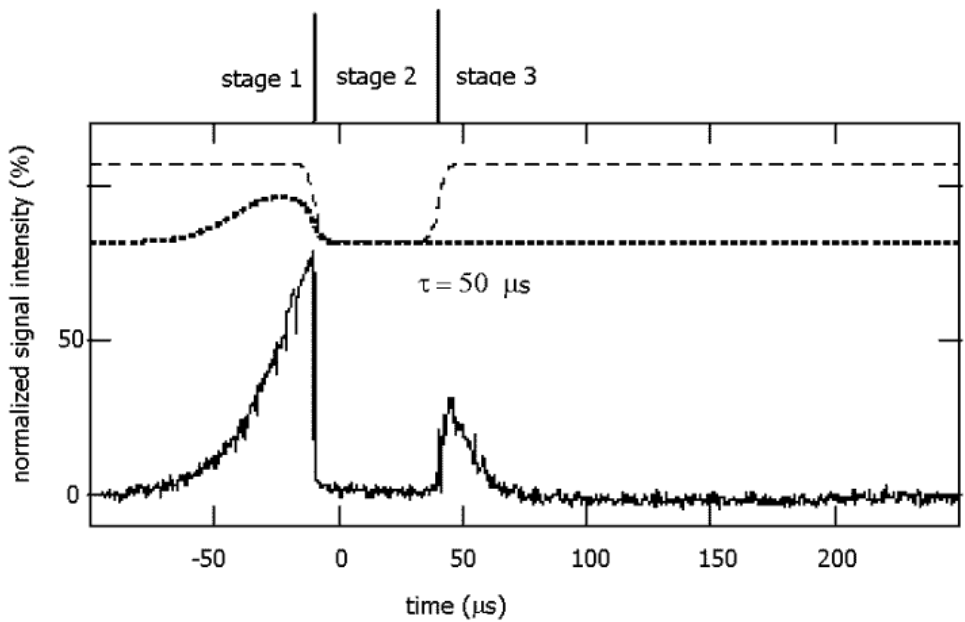}
\caption [OF]{A fragment of Fig. 2 from  \cite{phil} illustrating the
observed ``light pulse storage'' in a Rb vapor cell for a storage
time of 50 $\mu$s.  The curves in the upper part of the figure are
``calculated representations of the applied control field (dashed
line) and input signal pulse (dotted line)''.}
\end{figure}

\end{document}